\documentstyle[11pt,paspconf,epsf]{article}

\begin{document}

\title{Photometric Redshifts Of Starburst Galaxies}

\author{Charles T. Liu}
\affil{Department of Astronomy, Columbia University, Mail Code 5246,
    New York, NY 10027; and Department of Astrophysics, American Museum
    of Natural History, New York, NY  10024-5192}

\begin{abstract}
Although starburst galaxies have relatively flat spectral energy
distributions, their strong optical emission lines and near-UV continua
make it very feasible to estimate their redshifts photometrically.
In this work, I describe a photometric technique that simultaneously
{\bf (1)} identifies galaxies by star formation rate and 
{\bf (2)} measures their redshifts
with an accuracy of $\sigma_z = 0.05$ for objects at $z < 1$.
(An extension of the technique is potentially feasible, with the
use of near-infrared colors, to $1.6 < z < 2.5$.)
Applying this technique to a deep multicolor field survey
reveals a large excess population of strongly star-forming galaxies 
at $z \geq 0.3$ compared with $z < 0.3$.  Followup 
with spectroscopy and near-infrared photometry
confirms their presence, and suggests that some of them
may be in the midst of their initial burst of star formation.

\end{abstract}


\keywords{starburst galaxies, galaxy formation, photometric redshift techniques}

\section{Introduction}

The work presented here stems from a study of the field galaxy
population in the comparatively modest redshift range $z < 0.5$.
Using the photometric redshift method developed by Liu \& Green
(1998), Liu et al. (1998) found a substantial excess of relatively 
luminous ($M_B < -19$) starbursting galaxies at $z \geq 0.3$ compared 
with $z < 0.3$ in an optical multicolor survey of six deep blank 
sky fields (Hall et al. 1996).  First, I
discuss how and why the photometric technique can successfully identify
starbursts and estimate their redshifts.  I then present the results
of spectroscopic and near-infrared followup that verifies the
starburst excess, and hints at a surprising result: a significant
fraction of these galaxies may be undergoing their first major
star formation episode.

\section{Photometric Redshifts And Galaxy Spectral Types}

All photometric redshift techniques simultaneously
give at least some information about a galaxy's spectral type --
that is, its star formation rate per unit luminosity.  The method
of Liu \& Green (1998) seeks to optimize the ability to discrimminate
between spectral types without sacrificing redshift accuracy.
This is achieved by using a ``smallest maximum difference'' approach
comparing broad-band colors, rather than a ``least-squares'' approach
comparing broad-band fluxes.  (The distinction is roughly analogous
to using a Kolmogorov-Smirnov test rather than a $\chi^2$ test when
measuring the quality of a model fit to data.)  Another feature of this 
method is that it does not consider a galaxy's apparent
magnitude in the redshift determination; it is thus somewhat more
versatile when evaluating, for example, unusual or evolving
galaxy populations.

The method uses six broad-band filters (U/B/V/R/I7500/I8600)
and five completely empirical galaxy spectral energy
distributions (SEDs).  Shown in Figure 1, these SEDs 
represent objects with star formation rates typical
of E/S0, Sab, Sbc, Scd and Irr field galaxies. 
I refer the reader to Liu \& Green (1998) for further details;
the net result is an accuracy of $\sigma_z \sim 0.05$ in redshift
determination, and slightly better than $\pm 1$ in galaxy spectral type.

\begin{figure}
\plotfiddle{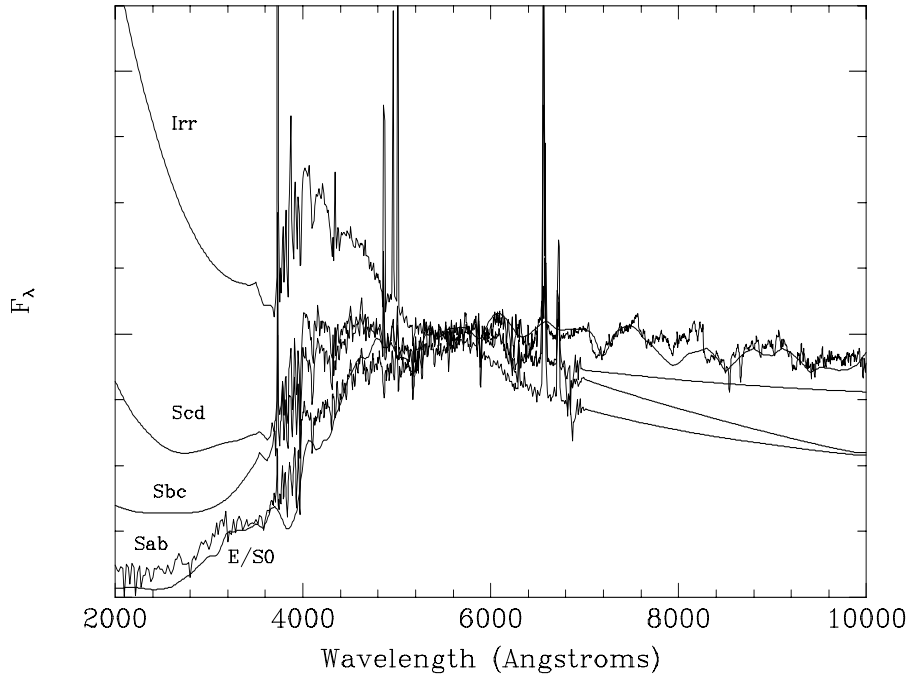}{3.5in}{0}{110}{130}{-342}{-378}
\caption{Spectral energy distributions of empirical galaxy templates,
in units of $F_{\lambda}$ normalized at 5500 \AA , from
Liu \& Green (1998).} \label{fig-1}
\end{figure}

Intuitively, it may seem that very blue galaxies such as starbursts
would present difficulties for photometric redshift determination.
After all, they have weak or no 4000 \AA\ breaks -- the most important
spectral feature for most photometric redshift
schemes -- and SEDs that are almost 
flat in $F_{\nu}$.  However, a glance at Figure 1 shows clearly that 
starburst SEDs have a steeply rising UV continuum blueward of 3500 \AA ,
a small but significant continuum ``hump'' around 4200 \AA ,
and very strong optical emission lines.  Together, these
characteristics make photometric redshifts of starbursts surprisingly
feasible.  Indeed,
the accuracy for starbursts using the smallest maximum difference
method is $\sigma_z \sim 0.05$, identical to the galaxy population 
as a whole.

\begin{figure}
\plotone{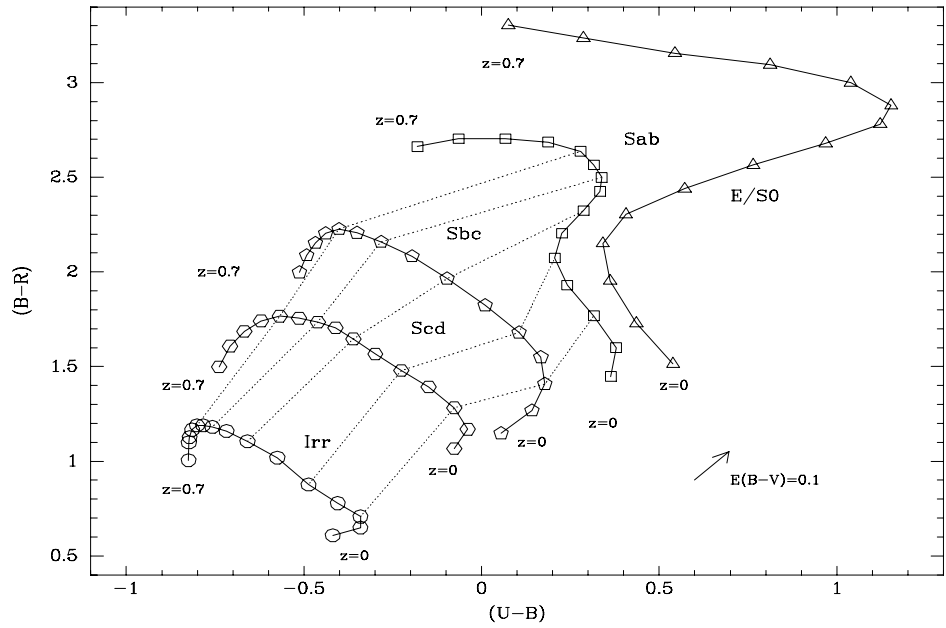}
\caption{Color evolutionary tracks in $U-B$ vs. $B-R$
for the empirically derived
template galaxy spectral types.  The tracks assume no luminosity
evolution with redshift.  Each point on the tracks represents a 
stepwise increase in $z$ of 0.05.  The dotted lines show the
iso-$z$ contours for $z = $ 0.1, 0.2, 0.3, 0.4 and 0.5.} \label{fig-2}
\end{figure}

Another way to look at it is to look in color-color space, as 
shown in Figure 2.  The solid lines show how the $U-B$ and $B-R$ colors
of the five galaxy SEDs change as a function of galaxy redshift, while
the dotted lines represent the ``iso-redshift contours'' 
at $z = $0.1, 0.2, 0.3, 0.4 and 0.5 for the galaxy SEDs
that have ongoing star formation. 
For these two colors, $0.1 < z < 0.5$ is the redshift range in which the
rest-frame near-UV continuum, the 4200 \AA\ hump,
and at least one of the three major rest-frame optical emission
line regions ([O\thinspace II], H$\beta + $[O\thinspace III], and
H$\alpha + $[N\thinspace II]) 
are all covered.  As a result, the iso-redshift
contours are roughly parallel and cleanly separated from each other.
This means that even very blue starburst galaxies can 
yield reliable photometric redshifts.  

Figure 2 reveals two more interesting points.  First, the effect of
dust extinction is shown by the vector in the lower right; it will move
any given galaxy SED roughly parallel to the iso-redshift contours.  So
a dusty galaxy may mimic a spectral type with a lower
star formation rate, but to first order its photometrically determined
redshift will not be strongly affected.  Second, these $UBR$ colors alone
yield surprisingly accurate photometric redshifts for strongly star-forming
galaxies ($\sigma_z \sim 0.07$).  Note that the $I$, $J$ and $H$ bands
cover in the range $1.6 < z < 2.5$ almost 
exactly the same rest wavelengths as the $U$, $B$ and $R$ bands cover 
in the range $0.1 < z < 0.5$ -- a happy coincidence!  Potentially,
IJH colors could be used to obtain photometric redshifts of galaxies,
with an expected accuracy of $\sigma_z \sim 0.20$ over the range
$1.6 < z < 2.5$, a somewhat difficult range in which to measure 
galaxy redshifts using optical spectra. This would then allow 
the direct comparison of similarly selected samples of
high, medium and low redshift star-forming galaxies.

\begin{figure}
\plotfiddle{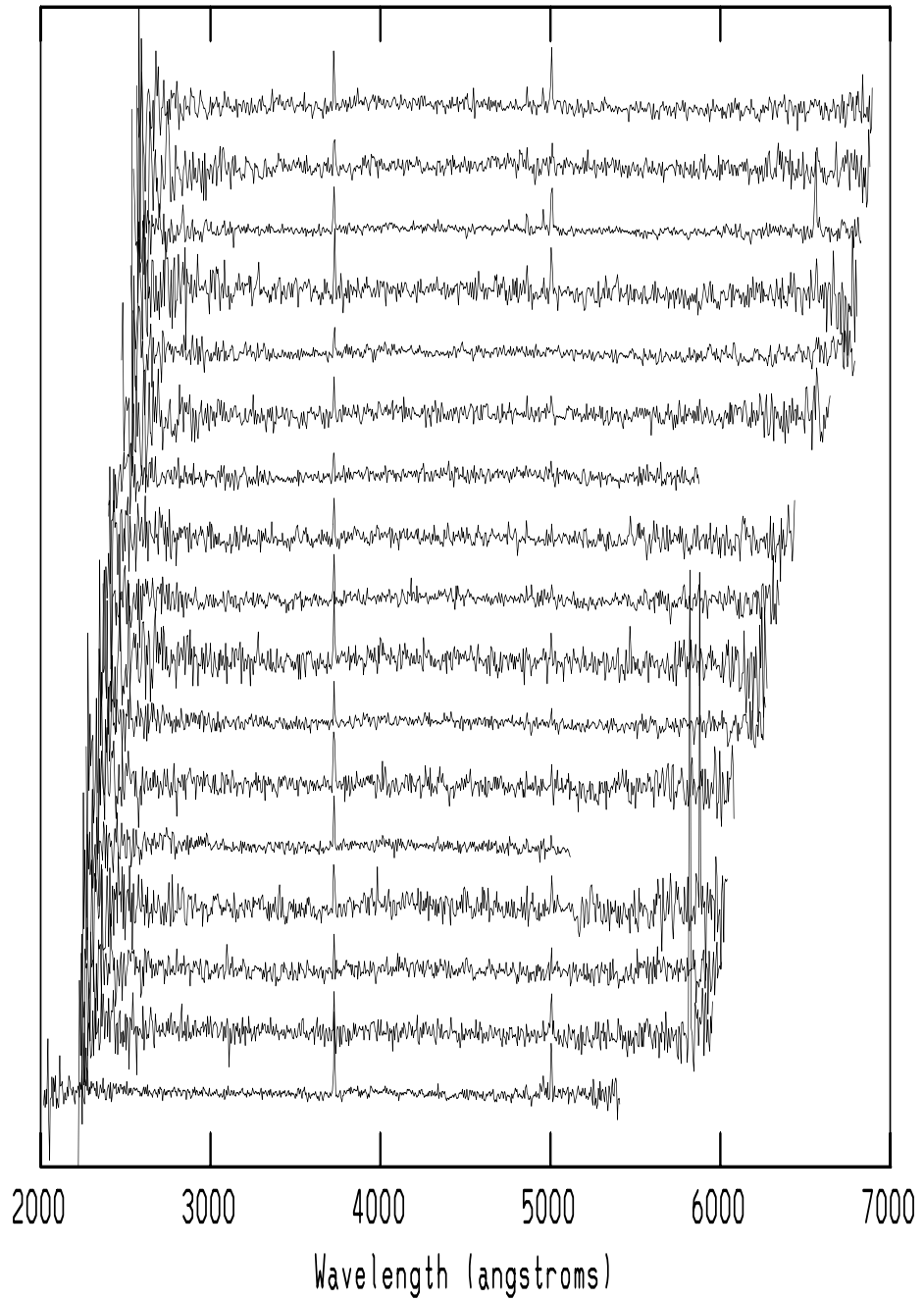}{6.5in}{0}{90}{180}{-279}{-468}
\caption{Spectra of 17 starburst galaxies representative of those
photometrically identified by Liu et al. (1998).  
The galaxies range from $z = 0.23$ (top) to
$z = 0.56$ (bottom), and have been deredshifted and plotted in 
arbitrary units of $F_{\lambda}$.  Note the strong 
[O\thinspace II]$\lambda$3727\AA\ emission line in every galaxy.
}
\end{figure}

\section{Spectroscopy And Infrared Photometry - \\
   Evidence For Galaxy Formation at $z \sim$ 0.3?}

The Deep Multicolor Survey of Hall et al. (1996), obtained at the
Mayall 4-meter telescope at KPNO, was the
testbed dataset for the Liu \& Green (1998) photometric redshift
method.  As described in Liu et al. (1998), a surprising excess of
starburst galaxies was detected in the range $0.3 \leq z \leq 0.5$.

As impressive as photometric redshifts may be, they are nonetheless only
statistically reliable.  For any given object, true confirmation
of its redshift must still come via spectroscopy.  With this in mind,
spectra were obtained of 30 of these objects, using the Steward Observatory
2.3-m telescope on Kitt Peak
and the Multiple Mirror Telescope on Mt. Hopkins.  Every single one
of these galaxies was confirmed to be a strongly star-forming galaxy.
Their redshifts range from $0.23 \leq z \leq 0.56$, and the rms error
of the photometric redshifts is 0.046, exactly consistent with prediction.
The rest-frame eequivalent widths of the 
[O\thinspace II]$\lambda$3727\AA\ emission line
range from 27 to 93 \AA , with a median of 40 \AA ; compare this with
a typical Sc galaxy, which has a type EW[O\thinspace II] of 10-20 \AA .
A representative sample of the galaxy spectra is presented in Figure 3.

Now the question is: what are these excess starbursting objects?
Two possibilities come to mind.  They could be galaxies temporarily
brightened by global starbursts, perhaps triggered by a merger event;
or they could be new galaxies undergoing their first major burst of
star formation.  In either scenario, such a galaxy would 
subsequently fade and redden at lower redshifts, disappearing into the
throng of ``ordinary'' field galaxies.  The way to distinguish between
these two cases is with infrared photometry, followed by differential
spectral synthesis to measure the mass of old stars in these objects.
The infrared observations are in progress, and so far H-band data have
been obtained for 21 of these starbursts.  (Alice Quillen of Steward 
Observatory is my collaborator in this endeavor.)
Preliminary results suggest that as many as half of them may have at
most 20\% of their mass in stars older than $10^8$ years.  Is this
possible evidence of new galaxies forming at $z \sim 0.3$?

Although this may be a tempting conclusion, I dare not draw it until 
more infrared data are obtained.  At present, these figures are reliable 
only at the 2-3$\sigma$ level, and any further claims would be purely
speculative.  At the redshifts of these objects, K-band data will be
absolutely necessary to quantify the presence of an old stellar
population.  I will therefore defer any further discussion of this
tantalizing idea until those data are in hand.


\acknowledgments

I thank the organizing committee for inviting me to this workshop,
and gratefully acknowledge support from NSF grant 
AST96-17177 to Columbia University.


%

\end{document}